\documentclass[12pt]{iopart}
\usepackage{iopams}
\usepackage{setstack}
\usepackage{amstext}
\usepackage{graphicx}
\usepackage{subfigure}
\usepackage{color}

\newcommand{\bra}[1]{\left\langle#1\right\vert}
\newcommand{\ket}[1]{\left\vert#1\right\rangle}
\newcommand{\abs}[1]{ \left\vert#1 \right\vert}

\begin{document}

\title[A quantum speedup in machine learning]{A quantum speedup in machine learning: Finding a $N$-bit Boolean function for a classification}

\author{Seokwon Yoo$^{1,2}$, Jeongho Bang$^{2,1}$, Changhyoup Lee$^{3,1}$, and Jinhyoung Lee$^{1,2}$}
\address{$^1$ Department of Physics, Hanyang University, Seoul 133-791, Korea}
\address{$^2$ Center for Macroscopic Quantum Control \& Department of Physics and Astronomy, Seoul National University, Seoul, 151-747, Korea}
\address{$^3$ Centre for Quantum Technologies, National University of Singapore, 3 Science Drive 2, Singapore 117543}
\ead{hyoung@hanyang.ac.kr}

\begin{abstract}
We compare quantum and classical machines designed for learning an N-bit Boolean function in order to address how a quantum system improves the machine learning behavior. 
The machines of the two types consist of the same number of operations and control parameters, but only the quantum machines utilize the quantum coherence naturally induced by unitary operators. 
We show that quantum superposition enables quantum learning that is faster than classical learning by expanding the approximate solution regions, i.e., the acceptable regions. 
This is also demonstrated by means of numerical simulations with a standard feedback model, namely random search, and a practical model, namely differential evolution.
\end{abstract}

\vspace{2pc}

\maketitle

%%%%%%%%%%%%%%%%%%%%%%%%%%%%%%%%%%%%%%%%%%%%%%%%%%%%%%%%%%%%
\section{Introduction}\label{sec:introduction}
%%%%%%%%%%%%%%%%%%%%%%%%%%%%%%%%%%%%%%%%%%%%%%%%%%%%%%%%%%%%
Over the past decades, quantum physics has brought remarkable innovations in the various fields of disciplines. 
For example, there are exponentially fast quantum algorithms compared to their classical counterparts \cite{Deutsch:1992dr,Grover:1997ht,Shor:1997bl}. 
The physical limit of measurement precision has been improved in quantum metrology \cite{Giovannetti:2004jg,Giovannetti:2011jk}, and a plenty of protocols offering higher security have been proposed in quantum cryptography \cite{Bennett:1984wv,Ekert:1991kl}. These achievements are enabled by appropriate usage of quantum effects such as quantum superposition and quantum entanglement. 

Another phenomenal science is machine learning which is a sub-field of artificial intelligence and one of the most advanced automatic control techniques.  
While learning is usually regarded as the character of humans or living things, machine learning enables a machine to learn a task \cite{Langley:1996vj}. 
Machine learning has been attracting great attention with its novel ability to learn. 
On one hand, machine learning has been studied for the understanding the learning of a real biological system, in a theoretical manner. 
On the other hand, it is also expected that machine learning provides reliable control techniques in designing the complex systems in a practical manner \cite{Langley:1996vj}. 

Recently, hybridizing two scientific fields described above, quantum technology and machine learning, has received great interest \cite{Pearson:2001ja, Gammelmark:2009kg, Bang:2012hr, Briegel:2012eu}. 
One question naturally arises: 
Can machine learning be improved by using favorable quantum effects? 
Several attempts to answer this question have been done in the past years, for example, quantum perceptrons \cite{Lewenstein:1994hs}, neural network \cite{Kak:1995ht,Chrisley:1996vi,Narayanan:2000hc}, quantum-inspired evolutionary algorithm \cite{Han:2002ku,Han:2006hd}. 
Most recently, remarkable studies have been made \cite{Manzano:2009gf, Bang:2013uj, Lloyd:2013us, Rebentrost:2013ux}. 
In Ref. \cite{Manzano:2009gf}, learning speedup of quantum machine was observed with less requirement of memory for a specific example, called $k$-th root NOT. 
In Ref. \cite{Bang:2013uj}, a strategy to design a quantum algorithm was introduced, establishing the link between the learning speedup and the speedup of the found quantum algorithm. 
In Refs. \cite{Lloyd:2013us, Rebentrost:2013ux}, the authors showed quantum speedup in the task of classifying large number of data. 
However, it is still unclear what and how quantum effects work in machine learning, particularly with the absence of fair comparison between classical and quantum machine.

In this work, we consider a binary classification problem as a learning task. 
Such classification can be realized to an $N$-bit Boolean function that maps a set of $N$-bit binary strings in $\{0,1\}^N$ into $\{0, 1\}$ \cite{Opper:1991wx}. 
The main work in this paper is to compare quantum machine with classical machine. 
These two machines are equivalent. 
The only differentiation is that the quantum machine can deal with quantum effects, whereas the classical machine cannot. 
The machines are analyzed in terms of acceptable region defined as a localized solution region of parameter space. 
In the analysis, it is shown that the quantum machine can learn faster due to the expanded acceptable region by quantum superposition. 
Such a quantum learning speedup is understood in terms of a expansion of the acceptable region. 
In order to make the analysis more explicit, we analyze further by using random search which is a standard model for the learning performance analysis \cite{Rastrigin:1963ua}. 
In such a primitive model, we validate the quantum speedup, showing that the overall number of iterations required to complete the learning is proportional to $O(\rme^{\alpha D})$, with $\alpha \simeq 3.065$ in classical machine, and $\alpha \simeq 0.238$ in quantum machine. 
Here, $D$ is the size of the search space. 
Differential evolution is employed as a learning model, taking into account more realistic circumstance.
By numerical simulations, we show that the quantum speedup is still observed even in such case.

%%%%%%%%%%%%%%%%%%%%%%%%%%%%%%%%%%%%%%%%%%%%%%%%%%%%%%%%%%%%
\section{Classical and quantum machines}\label{sec:machines}
%%%%%%%%%%%%%%%%%%%%%%%%%%%%%%%%%%%%%%%%%%%%%%%%%%%%%%%%%%%%
Machine learning can be decomposed into two parts, machine and feedback. 
Machine performs various tasks depending on its internal parameters, and feedback adjusts the parameters of machine for machine to perform a required task called target. 
Learning is a process finding suitable parameters of machine, whereby machine is expected to generates desired results for a target \footnote{
	We consider the case of supervised learning that desired results of task are given to the machine.
}. 
This concept of machine learning has been widely adapted in the context of machine learning at the fundamental level \cite{Langley:1996vj}.

In this work, we assign a machine a binary classification problem as a task, where machine will learn a target $N$-bit Boolean function, defined as
\begin{equation}
	f:\bi{x} \in \{ 0, 1\}^N \rightarrow y \in \{ 0, 1\}, 
	\label{EQ:booleanFunc} %EQ:booleanFunc
\end{equation}
where $\bi{x}=x_N \dots x_2 x_1$ is represented as a $N$-bit string of $x_j \in \{0, 1\}$ ($j=1,2,\ldots,N$). 
This function can be written by using Positive Polarity Reed-Muller expansion \cite{Gupta:2006ef},
\begin{equation}\fl
	f(\bi{x}) 
		= a_0\oplus a_1x_1 \oplus a_2x_2 \oplus a_3x_1x_2\oplus \cdots \oplus a_{2^N-1}x_1\cdots x_N 
		= \bigoplus_{k=0}^{2^N-1} 
			\left(
				a_k \prod_{j \in \mathrm{C}_k} x_j
			\right),
	\label{EQ:PPRM} %EQ:PPRM
\end{equation}
where $\oplus$ denotes modulo-$2$ addition, $\bigoplus$ means a direct sum of the modulus, and Reed-Muller coefficients $a_k$ are either $0$ or $1$. 
Here, $\mathrm{C}_k$ is an index set whose elements are given in such a way; 
The number $j$ is then taken to be an element of $\mathrm{C}_k$ only if $k_j$ is equal to $1$ when $k$ is written as a $N$-bit string, $k_N\dots k_2 k_1$.
Thus, each set of $\{ a_k \}$ corresponds to each of $2^{2^N}$ Boolean functions.

%Fig1_machine
\begin{figure}[tp]
	\centering
	\includegraphics[width=4.2in]{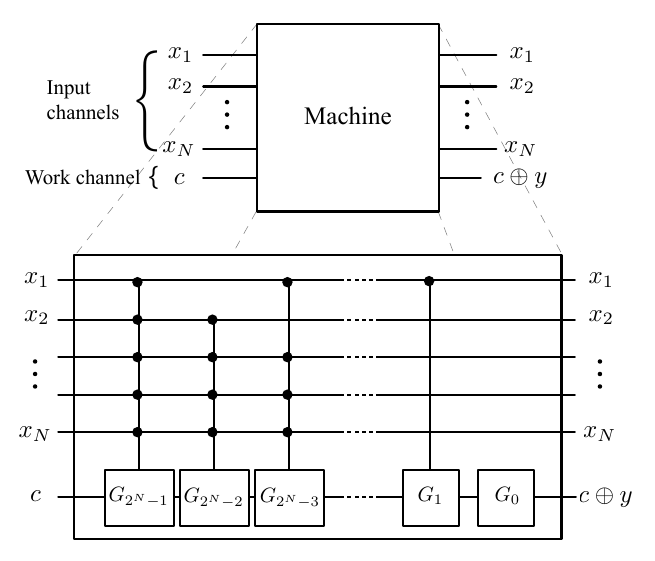}
	\caption{
		The N-bit Boolean function is implemented by a reversible circuit.
		The machine consists of $N$-bit input channels and single-bit work channel, which contains $2^N$ operations: 
		one single-bit operation $G_0$, and $2^N-1$ operations $G_{k}$ conditioned by input bits $\bi{x}$.
		Here, constant input $c$ is set to be $0$, which gives rise to output bit $y$.
	}
	\label{FIG:machine} %FIG:machine
\end{figure}
The Boolean function can be implemented by a reversible circuit as shown in \Fref{FIG:machine}, where an additional bit channel, called work channel, and controlled operations are employed \cite{Maslov:2003ux,Toffoli:1980tl}.
A single-bit operation $G_0$ is placed on the work channel and $(2^N-1)$ controlled-$G_k$ operations are acted on the work channel when all the control bits, $x_{j}$ ($j \in \mathrm{C}_k$), are $1$. 
The input signal $c$ on the work channel is fixed to $0$. 
The operation $G_{k}$ is given to be either identity (i.e., doing nothing) if $a_k=0$ or NOT (i.e., flipping an input bit to its complement bit) if $a_k=1$. 
As an example, $1$-bit Boolean function (i.e., $N=1$) has $2^{2^{1}}=4$ sets of Reed-Muller coefficients ($a_{0}$, $a_{1}$), which determine all possible Boolean functions. 
\Tref{TAB:1bitBoolean} gives four possible $1$-bit Boolean functions with Reed-Muller coefficients and corresponding operations.
%Table1
\begin{table}[h]
	\centering
	\tabcolsep=0.15in
	\begin{tabular}{c | c c | c c}
		\hline \hline
		Boolean function 		& $a_0$	 & $a_1$	 & ${G}_0$		& ${G}_1$	\\
		\hline
		$f_1:x \mapsto 0$ 		& $0$	 & $0$	 & Identity			& Identity \\
		$f_2:x \mapsto 1$ 		& $1$	 & $0$	 & NOT			& Identity \\
		$f_3:x \mapsto x$ 		& $0$	 & $1$	 & Identity			& NOT \\
		$f_4:x \mapsto x\oplus1$	& $1$	 & $1$	 & NOT			& NOT \\
		\hline \hline
	\end{tabular}
	\caption{
		Four possible $1$-bit Boolean functions are given with Reed-Muller coefficients ($a_{0}$ and $a_{1}$), and operations ($G_{0}$ and $G_{1}$). 
		These are common for both classical and quantum.
	}
	\label{TAB:1bitBoolean} %TAB:1bitBoolean
\end{table}

With a reversible circuit model, we then define classical and quantum machines. 
Classical machine consists of classical channels and operations, and the Boolean function of classical machine is described as
\begin{equation}
	\left(\bi{x}, c\right) \overset{f}{\longrightarrow} \left(\bi{x}, c \oplus y \right)
	\label{EQ:classicalMachine} %EQ:classicalMachine
\end{equation}
with classical bits $\bi{x}$, $y$, and $c$. 
We suppose that Reed-Muller coefficients $a_k$'s are probabilistically determined by internal parameters $p_k$'s, which implies $G_k$ performs identity and NOT operation with probabilities $p_k$ and $1-p_k$, respectively.
This probabilistic operations are primarily intended for a fair comparison with the quantum machine that naturally employs a probabilistic operation. 
Now, we construct quantum machine by setting only work channel to be quantum. 
The input channels are left in classical, as the input information is classical in our work. 
Thus, the Boolean function of quantum machine is described as
\begin{equation}
	\left(\bi{x}, \ket{c}\right) \overset{f}{\longrightarrow} \left(\bi{x}, \ket{\psi}\right),
	\label{EQ:quantumMachine} %EQ:quantumMachine
\end{equation}
where the signal on the work channel is encoded into a qubit state. 
The classical probabilistic operations $G_k$ are also necessarily replaced to unitary operators,
\begin{equation}
	\hat{G}_k =
		\left(
			\begin{array}{cc}
				\sqrt{p_k} 				& \rme^{\rmi \phi_k} \sqrt{1-p_k} \\
				\rme^{-\rmi \phi_k} \sqrt{1-p_k} 	& -\sqrt{p_k} \\
			\end{array}
		\right),
	\label{EQ:quantumUnitary} %EQ:quantumUnitary
\end{equation}
where $p_k$ is the probability of $\hat{G}_k$ performing identity, i.e., $\ket{0}\rightarrow\ket{0}$, $\ket{1}\rightarrow \rme^{\rmi\pi}\ket{1}$, and $1-p_k$ is that of $\hat{G}_k$ performing NOT, i.e., $\ket{0}\rightarrow\rme^{-\rmi\phi_k}\ket{1}$, $\ket{1}\rightarrow \rme^{\rmi\phi_k}\ket{0}$). 
Note that the relative phases $\phi_k$ are free parameters suitably chosen before the learning. 
The feedback adjusts only $p_k$'s, controllable both in classical and quantum experimental setups \cite{Reck:1994dz,Kim:2000bb}. 

These classical and quantum machines are equivalent each other. 
They have the same structures of circuit and the exactly same number of control parameters, $p_{k}$'s. 
Moreover, single classical operation $G_{k}$ and the quantum operator $\hat{G}_{k}$ cannot be discriminated by measuring distribution of outcomes for the same input $\bi{x}$ and $p_{k}$'s.

%%%%%%%%%%%%%%%%%%%%%%%%%%%%%%%%%%%%%%%%%%%%%%%%%%%%%%%%%%%%
\section{Acceptable region}\label{sec:ar}
%%%%%%%%%%%%%%%%%%%%%%%%%%%%%%%%%%%%%%%%%%%%%%%%%%%%%%%%%%%%
A target Boolean function is represented by a point, $Q_{f}=(p_0, p_1, \ldots, p_{2^N-1})$, in $2^N$-dimensional search space spanned by the probabilities, $p_k$'s. 
For example, four possible learning targets, $f_j$ ($j = 1,2,3,4$), of $1$-bit Boolean function correspond to four points on the search space; $\mathrm{Q}_{f_1}=(1,1)$, $\mathrm{Q}_{f_2}=(0,1)$, $\mathrm{Q}_{f_3}=(1, 0)$, and $\mathrm{Q}_{f_4}=(0, 0)$. 
Similarly, the machine behavior is also characterised as a point $\mathrm{Q}_\mathrm{m}=(p_0, p_1)$, i.e., the respective points lead to different probabilistic tasks that the machine performs. 
A learning is simply regarded as a process of moving $\mathrm{Q}_\mathrm{m}$ to a given target point in the whole search space. 
It is however usually impractical (actually, impossible in a realistic circumstance) to locate $\mathrm{Q}_\mathrm{m}$ exactly at the target point. 
Instead, it is feasible to find approximate solutions near to the exact target, i.e. the learning is expected to lead the point $\mathrm{Q}_\mathrm{m}$ into a region near to the target point \cite{Langley:1996vj}. 
We call such region acceptable region for the approximate target functions. 
As learning-time and convergence depend primarily on the size of the acceptable region, it is usually expected that larger acceptable region makes the learning faster \cite{Storn:1997hv}. 
In this sense, we examine the acceptable regions of classical and quantum machines. 

The acceptable region is defined as a set of points which guarantee the errors, $\epsilon = 1-{\cal F}$, less than or equal to a tolerable value, $\epsilon_t$. 
Here, ${\cal F}$ is the figure of merit of machine performance, called task-fidelity, to quantify how well the machine perform a target function, defined by
\begin{equation}
	{\cal F}(p_0, p_1, \ldots, p_{2^N-1}) = \left( \prod_\bi{x} \sum_y \sqrt{P(y|\bi{x}) P_\tau(y|\bi{x})} \right) ^{\frac{1}{2^N}},
	\label{EQ:taskFidelity} %EQ:taskFidelity
\end{equation}
where $P(y|\bi{x})$ is a conditional probability of obtaining an output $y$, given an input $\bi{x}$, and target probabilities $P_\tau(y|\bi{x})$ is that of the target. 
For example, we have a target probabilities, for $f_1$ in \Tref{TAB:1bitBoolean}, as
\begin{equation}
	P_\tau(0|0)=1,~ P_\tau(1|0)=0,~ P_\tau(0|1)=1,~\text{and}~ P_\tau(1|1)=0.
	\label{EQ:Ptarget} %EQ:Ptarget
\end{equation}
The term $\sum_y \sqrt{P(y|\bi{x}) P_\tau(y|\bi{x})}$ in \Eref{EQ:taskFidelity} corresponds to a closeness of the two probability distributions $P(y|\bi{x})$ and $P_\tau(y|\bi{x})$ for the given $\bi{x}$ \cite{Nielsen:2010vn}. 
The task fidelity, ${\cal F}$, increases as outputs get close to the required outputs; 
${\cal F}$ becomes unity only when the machine gives target for all $\bi{x}$, and otherwise, less than $1$.
The acceptable region can be seen as a set of probabilities, $p_{k}$'s, such that $1-\epsilon_{t} \le {\cal F}(p_{1},\cdots,p_{2^{N}-1})$, and thus, higher ${\cal F}$ guarantees a wider acceptable region for a given tolerance, $\epsilon_{t}$.

As the simplest case, let us begin with target function $f_1$\footnote{Such constant function, $f_1$, is one of trivial function, however, it is considerable for the machines to learn $f_1$.} of $1$-bit Boolean function, whose task fidelity, ${\cal F}(p_0, p_1)$, is reduced as
\begin{equation}
	{\cal F}(p_0, p_1) = \sqrt[4]{ P(0|0)P(0|1)},
	\label{EQ:Freduced1Bit} %EQ:Freduced1Bit
\end{equation}
which is common in both classical and quantum machines. 
In the classical machine, \Eref{EQ:Freduced1Bit} is evaluated as 
\begin{equation}
	{\cal F}_\mathrm{c}(p_0, p_1) = \sqrt[4]{p_0 ( p_0 p_1 + q_0 q_1 ) }, 
	\label{EQ:Fc} %EQ:Fc
\end{equation}
adopting the conditional probabilities $P_\mathrm{c}(y|\bi{x})$ given by
\begin{equation}
P_\mathrm{c}(0|0) = p_0 p_1 + p_0 q_1 = p_0, ~ P_\mathrm{c}(0|1) = p_0 p_1 + q_0 q_1, 
\end{equation}
where $q_j = 1 - p_j$ ($j=0,1$). 
In the quantum machine, the conditional probabilities $P_\mathrm{q}(y|\bi{x})$ slightly differ from $P_\mathrm{c}(y|\bi{x})$ due to the superposition between $\hat{G}_0$ and $\hat{G}_1$. 
The conditional probabilities $P_\mathrm{q}(y|\bi{x})$ are given as 
\begin{equation}
	\begin{array}{lll}
		P_\mathrm{q}(0|0) &=& \abs{ \bra{0}\hat{G}_0\ket{0} }^2 = P_\mathrm{c}(0|0), \nonumber \\
		P_\mathrm{q}(0|1) &=& \abs{ \bra{0}\hat{G}_1\hat{G}_0\ket{0} }^2 = P_\mathrm{c}(0|1) + p_\mathrm{int} \cos\Delta,
	\end{array}
	\label{EQ:Pquantum} %EQ:Pquantum
\end{equation}
where $p_\mathrm{int}=2\sqrt{p_0 p_1 q_0 q_1}$, and $\Delta = \phi_1 - \phi_0$ is difference of the phases in the two unitaries $\hat{G}_0$ and $\hat{G}_1$. 
Thus, the task-fidelity ${\cal F}_\mathrm{q}$ of quantum machine is evaluated as
\begin{equation}
	{\cal F}_\mathrm{q}(p_0, p_1) = \sqrt[4]{{\cal F}_\mathrm{c}^4 + p_0 p_\mathrm{int} \cos \Delta},
	\label{EQ:Fq} %EQ:Fq
\end{equation}
where the additional term of $\cos\Delta$ is apparently the result of quantum superposition. 
From the result of \Eref{EQ:Fq}, we can see that
\begin{equation}
	\left\{
		\begin{array}{cc}
			{\cal F}_\mathrm{q} > {\cal F}_\mathrm{c}	&~~\text{if } \cos\Delta > 0, \\
			{\cal F}_\mathrm{q} < {\cal F}_\mathrm{c}	&~~\text{if } \cos\Delta < 0, 
		\end{array}
	\right. 
	\label{EQ:interference} %EQ:interference
\end{equation}
provided that $0<p_{j}<0$ ($j=0,1$).
The phase $\Delta$ plays an important role in helping quantum machine by constructive interference leading to ${\cal F}_\mathrm{q}>{\cal F}_\mathrm{c}$. 
The task fidelities for other three targets are also listed in \Tref{TAB:taskFidelity}. 
%Table2
\begin{table}[h]
	\centering
	\tabcolsep=0.2in
	\begin{tabular}{ c | c c}
		\hline \hline
		Function & ${\cal F}_\mathrm{c}(p_0, p_1)$ & ${\cal F}_\mathrm{q}(p_0, p_1)$ \\
		\hline
		$f_1$ & {$\sqrt[4]{p_0 ( p_0 p_1 + q_0 q_1 ) }$} & {$\sqrt[4]{{\cal F}_\mathrm{c}^4 + p_0 p_\mathrm{int} \cos \Delta }$} \\
		$f_2$ & {$\sqrt[4]{q_0 ( q_0 p_1 + p_0 q_1 ) }$} & {$\sqrt[4]{{\cal F}_\mathrm{c}^4 - q_0 p_\mathrm{int} \cos \Delta }$} \\
		$f_3$ & {$\sqrt[4]{p_0 ( q_0 p_1 + p_0 q_1 ) }$} & {$\sqrt[4]{{\cal F}_\mathrm{c}^4 - p_0 p_\mathrm{int} \cos \Delta }$} \\
		$f_4$ & {$\sqrt[4]{q_0 ( p_0 p_1 + q_0 q_1 ) }$} & {$\sqrt[4]{{\cal F}_\mathrm{c}^4 + q_0 p_\mathrm{int} \cos \Delta }$} \\
		\hline \hline
	\end{tabular}
	\caption{
		The task-fidelities of quantum and classical machines are given in terms of probabilities ($p_{0}$ and $p_{1}$) for each target function of 1-bit Boolean function. 
		The phase $\Delta$ is defined in the main text, which plays an important role in quantum machine learning. 
	}
	\label{TAB:taskFidelity} %TAB:taskFidelity
\end{table}
Note here that, for all cases of target function $f_j$, ${\cal F}_\mathrm{q}$ can always be larger than ${\cal F}_\mathrm{c}$ by choosing appropriate free parameters $\phi_1$ and $\phi_2$ before the learning. 
Therefore, the quantum machine has wider acceptable regions than the classical machine for a given tolerance. 
%Fig2_acceptableRegion
\begin{figure}[tp]
	\centering
	\subfigure[Classical machine]
		{\includegraphics[width=3.10in]{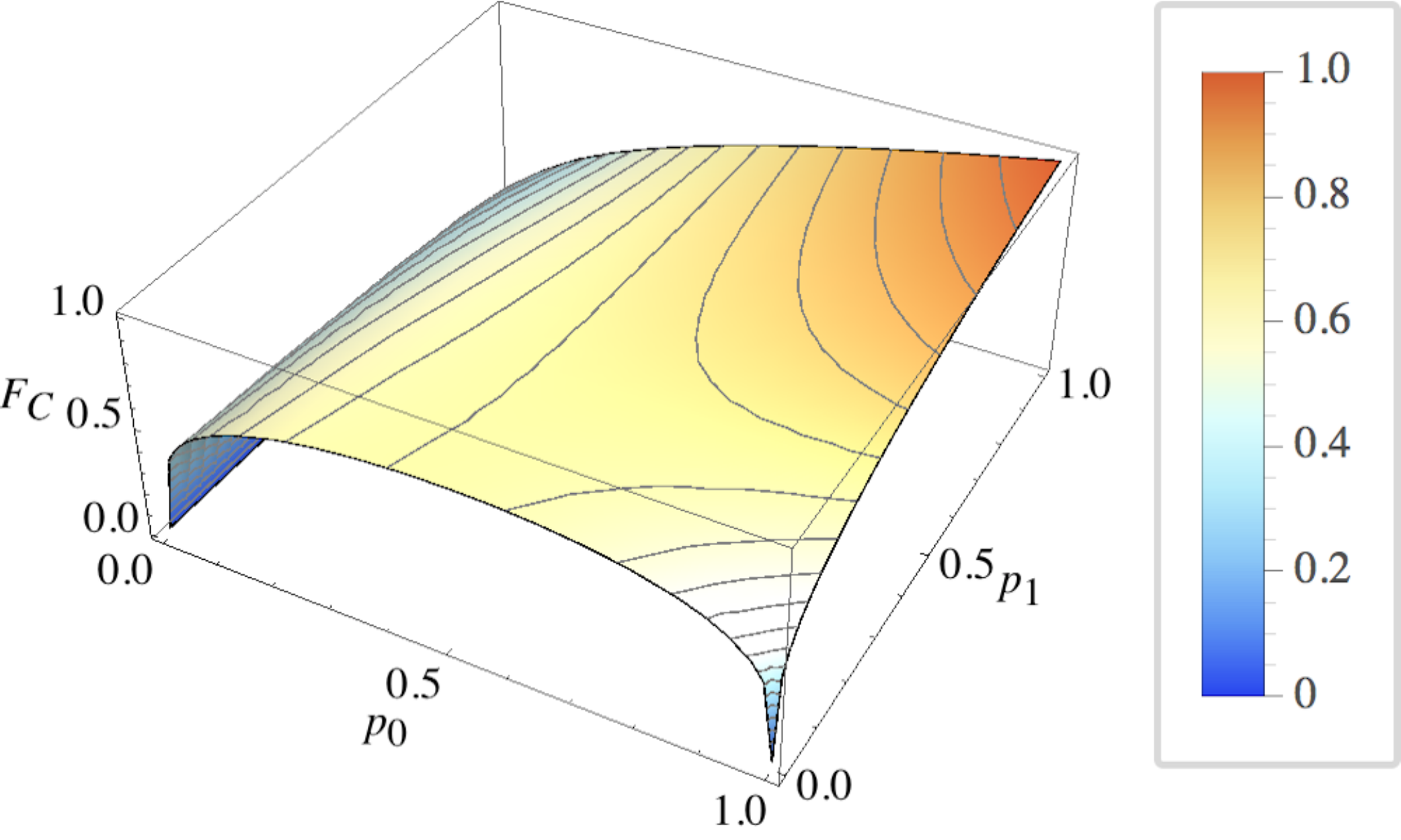}}
		{\includegraphics[width=2.48in]{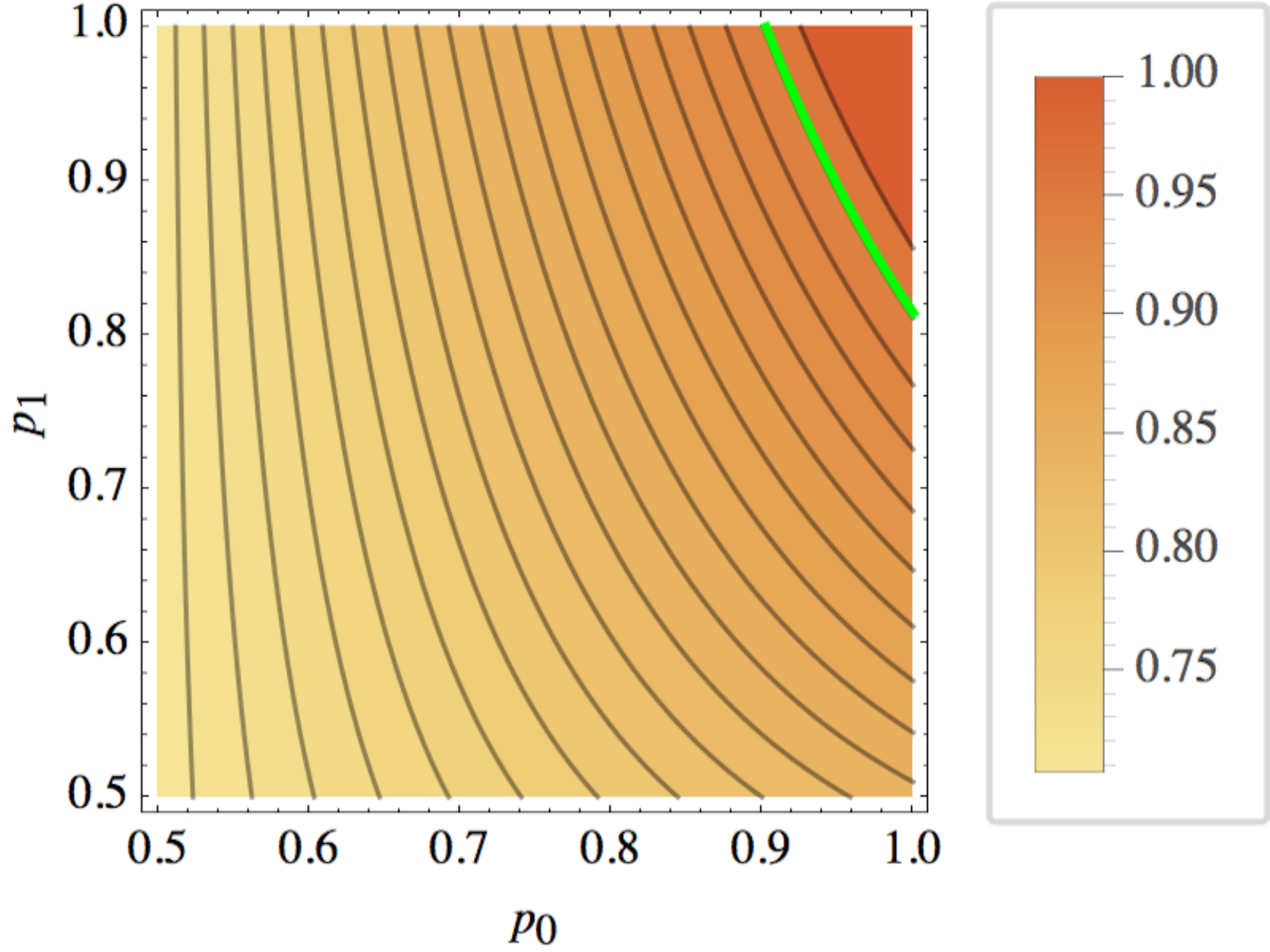}}
	\subfigure[Quantum machine]
		{\includegraphics[width=3.10in]{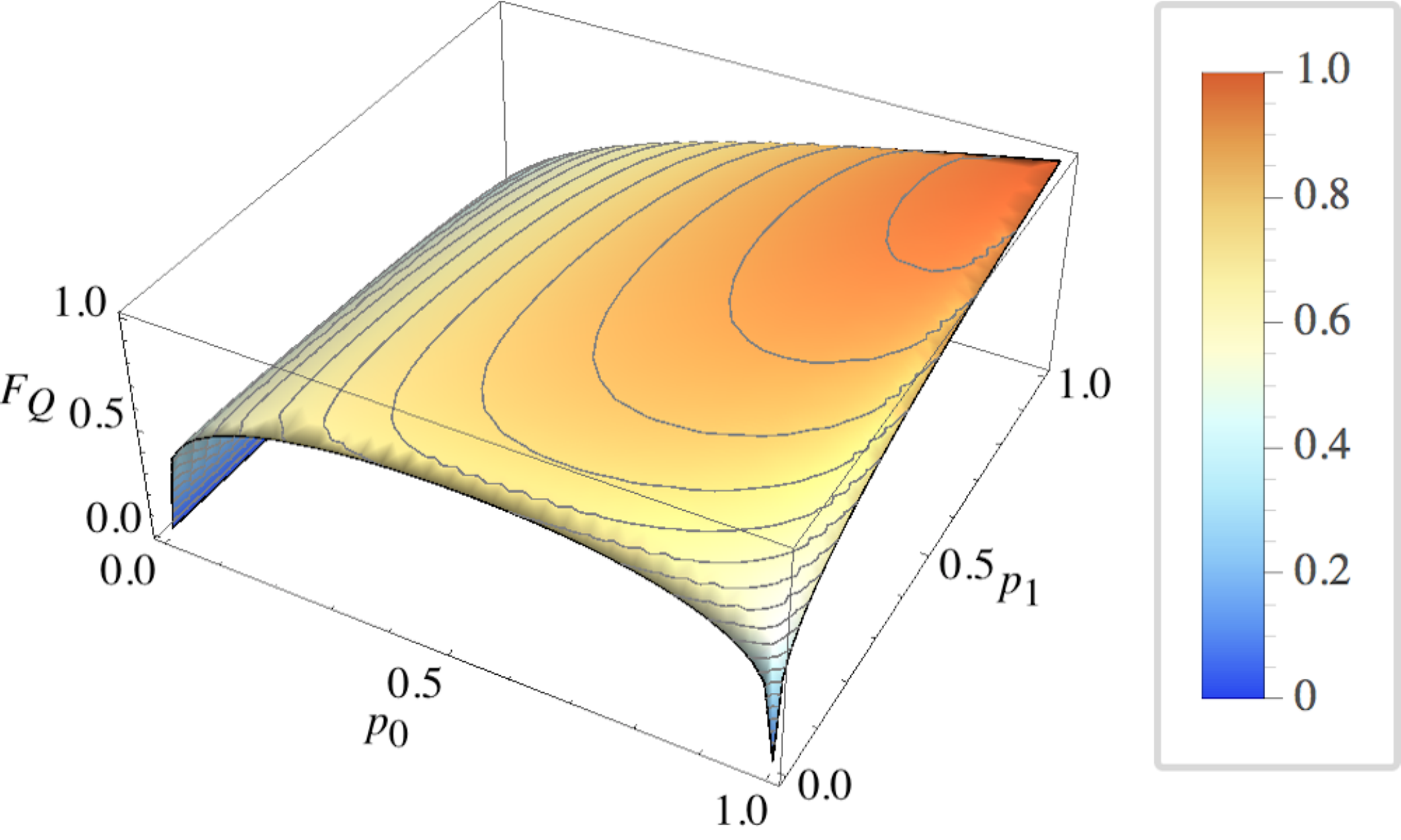}}
		{\includegraphics[width=2.48in]{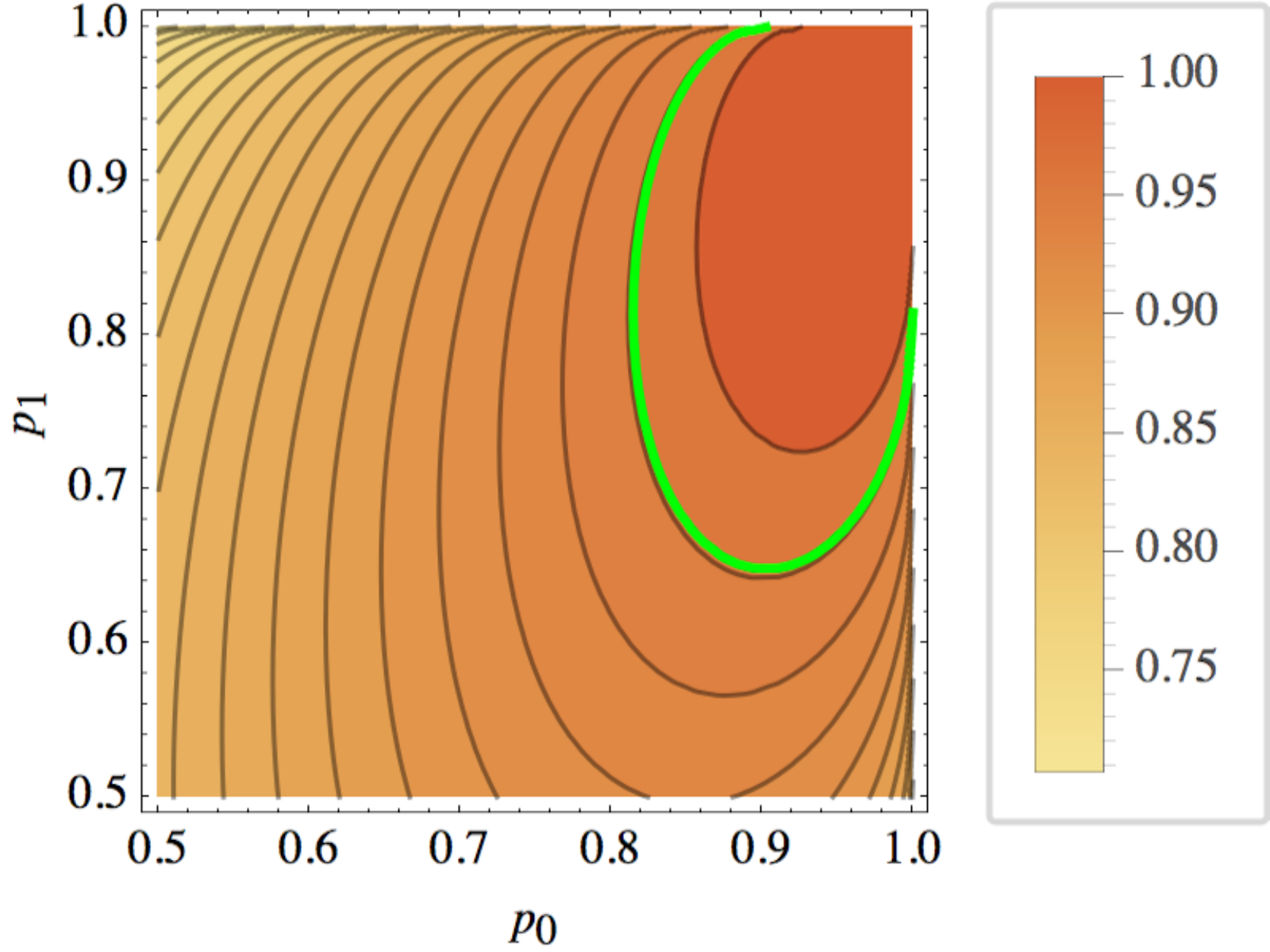}}
	\caption{
		Left column: the task fidelities for classical and quantum machines. 
		Right column: Green lines in the magnified views indicate the acceptable regions for a given tolerable error $\epsilon_t=0.05$ around the exact target point, $(p_{0},p_{1})=(1,1)$. 
		Here, we set $\Delta=0$ to maximize the task fidelity of quantum machine. It is found that the acceptable region of quantum machine is about $5.6$ times larger than that of classical machine.
	}
	\label{FIG:acceptableRegion} %FIG:acceptableRegion
\end{figure}
In \Fref{FIG:acceptableRegion}, the task-fidelity and the acceptable region for each machine are shown for the target $f_1$ when $\Delta = 0$ is chosen to maximize the difference between the two machines. 
We also found that the acceptable region of quantum machine is larger about $5.6$ times than that of classical machine.

The optimal phase condition to improve the task-fidelity, as in \Eref{EQ:interference}, can be generalized to arbitrary $N$-bit Boolean function ($N > 1$).
We provide one of the conditions as 
\begin{equation}
	\phi_k = 
	\left\{
		\begin{array}{cc}
			0	&~~\text{if } s_k = 0 \\
			\pi	&~~\text{if } s_k = 1
		\end{array}
	\right. .
	\label{EQ:optimizedPhase} %EQ:optimizedPhase
\end{equation}
where, $s_k$ is $k$th component of a solution point $\mathrm{Q}_f (s_0, s_1, \ldots, s_{2^N-1})$ in $2^N$ dimensional search space (See \ref{appendix_a}). 
This condition yields ${\cal F}_\mathrm{q}\ge{\cal F}_\mathrm{c}$, so that the acceptable region of quantum machine can be wider than classical machine for arbitrary $N$-bit Boolearn function.

%%%%%%%%%%%%%%%%%%%%%%%%%%%%%%%%%%%%%%%%%%%%%%%%%%%%%%%%%%%%
\section{Learning speedup by expanded acceptable region}\label{sec:lt}
%%%%%%%%%%%%%%%%%%%%%%%%%%%%%%%%%%%%%%%%%%%%%%%%%%%%%%%%%%%%
This section is devoted to learning-time in machine learning. 
For a numerical simulation, we employ random search as a feedback, which has been often considered for studying learning performance rather than for any practical reasons  \cite{Rastrigin:1963ua}.
Random search runs as follow: 
First, all $2^N$ control parameters $p_k$ are randomly chosen, and then, task-fidelity is measured with the chosen $p_k$'s. 
These two steps are thought of as a single iteration of the procedure. 
The iterations are repeated until the condition ${\cal F} \ge 1-\epsilon_t$ is satisfied for a given $\epsilon_t$.
After a sufficient number of simulations is performed, we then calculate the mean iteration number defined as  $n_{c}=\sum nP(n)$, where $P(n)$ is the probability to complete learning at the $n$th iterations. 
This mean iteration number, $n_c$, can be used to quantify the learning-time, and the results of numerical simulations for $n_{c}$ are shown in table 3, where quantum learning is demonstrated to be faster than classical learning. 
This is a direct result of the wider acceptable region of quantum machine as $n_{c}$ is inversely proportional to the size of acceptable region in random search; $n_{c}=1/\gamma$ is given by substituting $P(n)=\gamma(1-\gamma)^{(n-1)}$, where $\gamma$ is equal to the ratio of the acceptable region to the whole space in random search. 
We demonstrate this by comparing the results of $n_{c}$ with the acceptable regions $\gamma$ found by Monte-Carlo simulation in \Tref{TAB:gammaAndLearningTime}, and thereby we note the acceptable region is the main feature which directly influences learning-time in random search. 

\begin{figure}[tp]
	\centering
	\includegraphics[width=3.225in]{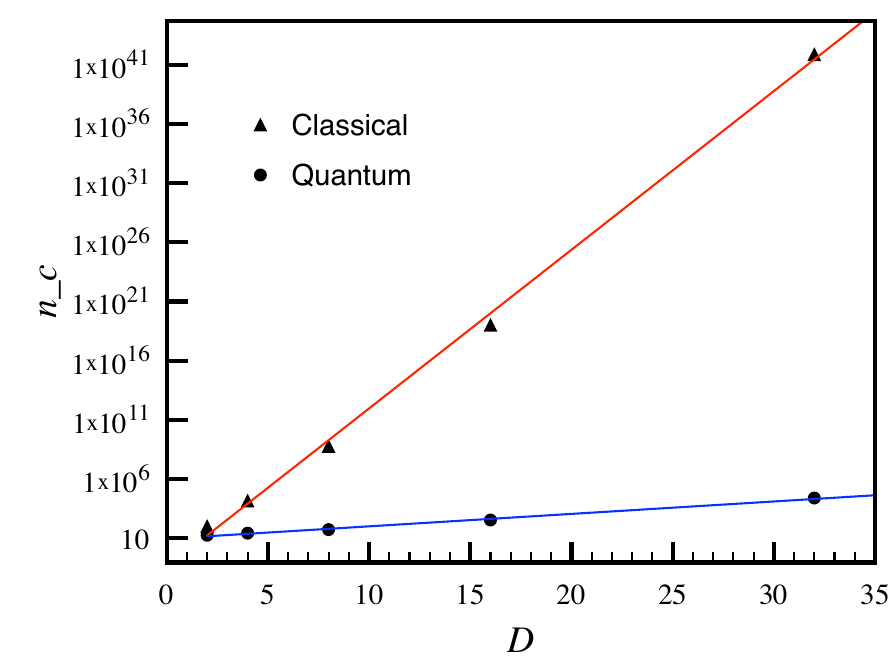}
	\caption{
		The learning-time, $n_{c}$, with dimension $D=2^N$ of parameter space for $1000$ realisations. 
		In this work, we consider a constant target function that yields $0$ for all input $\bi{x}$, the optimal phase condition of \Eref{EQ:optimizedPhase} is chosen for quantum machine, and the tolerable error $\epsilon_t$ is set to be $0.05$. 
		The data are well fitted to $\ln{n_c}=\alpha D + \beta$ in classical (red line) and quantum (blue line), with the fitting parameters $\alpha$ and $\beta$ as in \Eref{EQ:fittingParameter}.
	}
	\label{FIG:result_RS} %FIG:result_RS
\end{figure}

Also in \Fref{FIG:result_RS}, the data for $n_{c}$ in \Tref{TAB:gammaAndLearningTime} are well fitted to a function $\ln{n_c}=\alpha D + \beta$, implying that the size of the acceptable region is exponentially decreased as dimension $D=2^N$ of parameter space increases, i.e. $n_c = O(e^{\alpha D})$ \cite{vandenBergh:2004kz}. 
The fitting parameters are given as
\begin{equation}
	\left\{
		\begin{array}{ll}
			\alpha \simeq 3.065 \pm 0.072, ~\beta \simeq -3.188 \pm 1.196 &~\text{in classical case}, \\
			\alpha \simeq 0.238 \pm 0.008, ~\beta \simeq 2.267 \pm 0.127 &~\text{in quantum case}.
		\end{array}
	\right.
	\label{EQ:fittingParameter} %EQ:fittingParameter
\end{equation}
Remarkable is that the exponent $\alpha$ in quantum is much smaller than that in classical. 

%Table3
\begin{table}[h]
	\centering
	\tabcolsep=0.2in
	\begin{tabular}{ c | c c | c c}
		\hline \hline
		& \multicolumn{2}{c |}{Classical} & \multicolumn{2}{c}{Quantum}\\
		$N$ & $\gamma^{-1}$  & $n_c$ & $\gamma^{-1}$  & $n_c$ \\
		\hline
		$1$	& $1.0 \times 10^{2}$	& $1.03\times10^{2}$	& $1.8 \times 10^{1}$	& $1.74\times10^{1}$ \\
		$2$	& $1.4 \times 10^{4}$	& $1.39\times10^{4}$	& $2.6 \times 10^{1}$	& $2.68\times10^{1}$ \\
		$3$	& $4.4 \times 10^{8}$	& $4.67\times10^{8}$	& $5.5 \times 10^{1}$	& $5.36\times10^{1}$ \\
		$4$	& $9.8 \times 10^{18}$	& -					& $3.5 \times 10^{2}$	& $3.48\times10^{2}$ \\
		$5$	& $7.1 \times 10^{41}$	& -					& $2.5 \times 10^{4}$	& $2.48\times10^{4}$ \\
		\hline \hline
	\end{tabular}
	\caption{
		The learning-time $n_c$ is compared with the acceptable regions $\gamma$, where it is demonstrated that $n_c = \gamma^{-1}$. 
		This implies larger acceptable region leads less learning-time. 
		Simulation has failed for $N=4$ and $5$ in classical case due to finite computational resources for a very long run-time. 
	}
\label{TAB:gammaAndLearningTime} %TAB:gammaAndLearningTime
\end{table}

%Fig4_learning_time_RS

It follows from what has been shown that acceptable region is the main feature which directly influences learning-time in random search. 
We have proved that we can always prepare quantum machine which has larger acceptable region than classical one, in the previous chapter.
Therefore, we finally conclude that the learning-time can be shorter in quantum than in classical case. 
The result of numerical simulation also support that quantum machine learns much faster, particularly in a large search space. 
We clarify again that such a quantum speedup is enabled by the quantum superposition, and appropriately arranged phases.

%%%%%%%%%%%%%%%%%%%%%%%%%%%%%%%%%%%%%%%%%%%%%%%%%%%%%%%%%%%%
\section{Applying differential evolution}\label{sec:de}
%%%%%%%%%%%%%%%%%%%%%%%%%%%%%%%%%%%%%%%%%%%%%%%%%%%%%%%%%%%%
We consider more practical learning model, taking into account a real circumstance. 
A general analysis of the learning efficiency is very complicated as too many factors are associated with the learning behavior. 
Furthermore, the most efficient learning algorithms tend to use the heuristic rules and are problem-specific \cite{Middleton:2004ji,Pal:1996bm}. 
Nevertheless, it is usually believed that the acceptable region is a key factor of the learning efficiency in a heuristic manner \cite{vandenBergh:2004kz}. 
In this sense, we conjecture that the quantum machine offers the quantum speedup even in a practical learning method. 

We apply differential evolution (DE) which is known as one of the most efficient learning methods for the global optimization \cite{Storn:1997hv}. 
We start with $M$ sets of control parameter vectors $\bi{p}_i=(p_0, p_1, \dots, p_{2^N-1})_i$, for $i=1,2,\dots,M$, whose components are the control parameters of machine. 
In DE, these vectors, $\bi{p}_i$, are supposed to evolve by mating their components $p_k$'s with each other. 
\Eref{EQ:taskFidelity} is used as a criteria how well machines with $\bi{p}_i$ fit to the target. 
This process is iterated until the task-fidelity reaches a certain level of accuracy $1-\epsilon_t$ (See reference \cite{Storn:1997hv} or \cite{Bang:2013uj} for detailed method of the differential evolution).

\begin{figure}[tp]
	\centering
	\subfigure[average fidelity]
		{\includegraphics[width=3in]{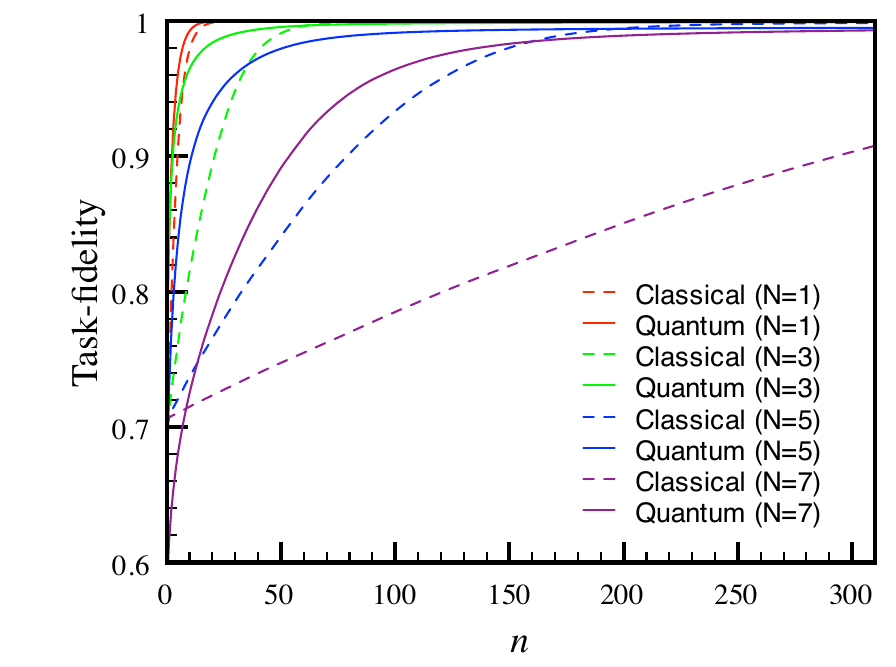}}
	\subfigure[learning time]
		{\includegraphics[width=3in]{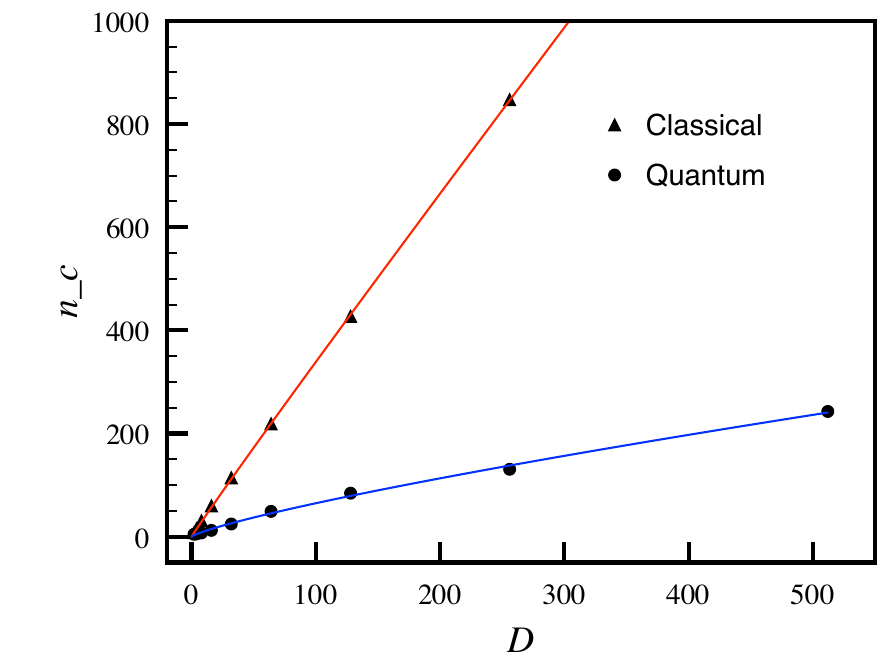}}
	\caption{
		(a) The mean task-fidelity are given with respect to the iteration $n$. 
		The simulations are done increasing $N$ from $1$ to $7$. 
		It is easily observed that the increments of the task-fidelities are faster in quantum for all cases. 
		(b) The learning time, $n_\mathrm{c}$, is drawn as dimension $D$ of parameter space increase.
		The data are well fitted to a presumable function $n_\mathrm{c} \simeq \alpha D^\beta$, with $\alpha \simeq 3.82$, $\beta \simeq 0.97$ in classical (red line), and $\alpha \simeq 1.61$, $\beta \simeq 0.80$ in quantum (blue line). 
		Note that the quantum machine still shows better convergence, with the smaller $\alpha$ and $\beta$.
	}
	\label{FIG:resultDE} %FIG:resultDE
\end{figure}

We perform the numerical simulations by increasing $N$ from $1$ to $7$. 
The results are averaged over $1000$ realisations for $M=50$ and $\epsilon_t = 0.05$.
The target function is a constant function, $f(\bi{x})=0$ for all $\bi{x}$. 
Free parameters in differential evolution (e.g., crossover rate and differential weight) are chosen to achieve the best learning efficiency for classical machine\footnote{
	In \Fref{FIG:resultDE}(a), one may worry about the crossover point (for $N \ge 5$) associated with validity of the quantum learning speedup for $\epsilon_t \rightarrow 0$. 
	However, the appearance of the crossover is due to the DE optimization with the free parameters. 
	Note here that the free parameters are optimized the classical machine. 
	The crossover can be removed by choosing the appropriate free parameters for each machine.
}. 
Nevertheless, we expect that the quantum machine still exhibits the quantum speedup, assisted by the quantum superposition, with the optimal phases in \Eref{EQ:optimizedPhase}. 
%Fig5_fidelity_DE, Fig5_learning_time_DE
We give the mean task-fidelity averaged over $M$, in \Fref{FIG:resultDE}(a). 
For both classical and quantum, the mean task-fidelities are increased close to $1$, but quantum machine is much faster for all cases. 
We investigate learning-time $n_c$ as increasing the dimension $D=2^N$ of parameter space, as depicted in \Fref{FIG:resultDE}(b). 
The data are well fitted to a presumable function $n_\mathrm{c} \simeq \alpha D^\beta$, with $\alpha \simeq 3.82$, $\beta \simeq 0.97$ in classical machine, and $\alpha \simeq 1.61$, $\beta \simeq 0.80$ in quantum machine \footnote{
	Such polynomial result shows much improvement from the differential evolution, which is quite distinct from the case of random search which exhibits the exponential dependence.
}. 
We note that the quantum machine still exhibits the speedup, with the smaller $\alpha$ and $\beta$.
Therefore, we expect that such quantum speedup can be achieved even in a real circumstance.

%%%%%%%%%%%%%%%%%%%%%%%%%%%%%%%%%%%%%%%%%%%%%%%%%%%%%%%%%%%%
\section{Summary and discussion}\label{sec:summary}
%%%%%%%%%%%%%%%%%%%%%%%%%%%%%%%%%%%%%%%%%%%%%%%%%%%%%%%%%%%%
We investigated learning performance of two machines by considering the task of finding a $N$-bit Boolean function which can be used in a binary classification problem. 
The two machines were equivalently designed to make the comparison of these two machine as convincing as possible. 
The critical difference between the two machines was that the operations in quantum machine are described by unitary operators to deal with the quantum superposition. 
The learning of the two machines were characterized in terms of acceptable region, the localized region of the parameter space including approximate solutions.
We have found that the quantum machine has a wider acceptable region, induced by quantum superposition. 
We demonstrated simulation with a standard feedback method, random search, to show that the size of acceptable regions were inversely proportional to the learning-time. 
Here, it was also shown that the wider acceptable region make the learning faster; namely, the learning-time is proportional to $O(\rme^{\alpha D})$, with $\alpha \simeq 3.065$ in the classical learning and $\alpha \simeq 0.238$ in the quantum machine. 
We then applied a practical learning method, differential evolution, to our main task, and observed the learning speedup of quantum machine.

Here, we would like to remind that the maximized learning speedup of the quantum machine is achieved by choosing the suitable phases as in \Eref{EQ:optimizedPhase}. 
From a practical perspective, one may consider that an additional task, such as finding the relative phases, is required to ensure remarkable performance of the quantum learning machine for other $N$-bit Boolean function targets. 
Alternatively, such an issue can be overcome by synchronizing the relative phases with the control parameters in the quantum machine, still yielding the learning speedup (see \ref{appendix_b} for details). 

We expect that our work motivates researchers to study the role of various quantum effects in machine learning, and open up new possibilities to improve machine learning performance. It is still open whether the quantum machine can be improved more by using other quantum effects, such as quantum entanglement.

%%%%%%%%%%%%%%%%%%%%%%%%%%%%%%%%%%%%%%%%%%%%%%%%%%%%%%%%%%%%
\section*{Acknowledgment}
%%%%%%%%%%%%%%%%%%%%%%%%%%%%%%%%%%%%%%%%%%%%%%%%%%%%%%%%%%%%
We acknowledge the financial support of the National Research Foundation of Korea (NRF) grant funded by the Korea government (MEST) (No. 2010-0018295 and No. 2010-0015059).
We also thanks T. Ralph, M. \.{Z}ukowski and H. J. Briegel for discussion and advice. 

%%%%%%%%%%%%%%%%%%%%%%%%%%%%%%%%%%%%%%%%%%%%%%%%%%%%%%%%%%%%
\appendix
%%%%%%%%%%%%%%%%%%%%%%%%%%%%%%%%%%%%%%%%%%%%%%%%%%%%%%%%%%%%
\section{Finding the optimal phase condition in \Eref{EQ:optimizedPhase}}\label{appendix_a}
%%%%%%%%%%%%%%%%%%%%%%%%%%%%%%%%%%%%%%%%%%%%%%%%%%%%%%%%%%%%
Let us recall the general form of task fidelity as in \Eref{EQ:taskFidelity}. 
We suppose the target as a deterministic function. 
Then, \Eref{EQ:taskFidelity} is rewritten as
\begin{equation}
	{\cal F}(p_0, p_1, \ldots, p_{2^N-1}) = \left( \prod_\bi{x} P(f(\bi{x})|\bi{x}) \right) ^{\frac{1}{2^{N+1}}}.
	\label{EQ:FreducedGeneral} %EQ:FreducedGeneral
\end{equation}
In deriving the above reduced form of \Eref{EQ:FreducedGeneral}, we used that $P_\tau(y|\bi{x})=1$ when $y$ is equal to the desired value $f(\bi{x})$ for a given target $f$, and otherwise $P_\tau(y|\bi{x})=0$. 
\Eref{EQ:FreducedGeneral} shows that the task fidelity is enlarged if $P(f(\bi{x})|\bi{x})$ for all $\bi{x} \ne 0$ are maximized.

To start, consider an ideal learning machine (either classical or quantum) that always generates the desired outcome results with perfect task-fidelity ${\cal F}=1$. 
From our analysis in \Sref{sec:ar}, we can represent this machine as a point $\mathrm{S} = (s_0, s_1, \dots , s_{2^N-1})$ in $2^N$-dimensional search space. In this sense, we call this ideal machine ``solution machine''. 
We then consider a ``near-solution machine'' which is located on a point $\mathrm{Q} = (p_0, p_1, \dots , p_{2^N-1})$ in the search space. 
More specifically, $d(\mathrm{Q},\mathrm{S}) = \sqrt{\sum_{k=0}^{2^N-1} (s_k - p_k)^2}=\delta$, where $d(\mathrm{Q},\mathrm{S})$ is Euclidean distance. 
Here we assume further that the search space is isotropic around $\mathrm{S}$ so that the machines on the surface of the hyper sphere $d(\mathrm{Q},\mathrm{S})=\delta$ have the same task-fidelity. 
This assumption is physically reasonable for very small tolerance error. 
Thus, without loss of generality, we consider the near-solution machine corresponding to the point $\mathrm{Q}$ on the sphere $d(\mathrm{Q},\mathrm{S})=\delta$, satisfying $\abs{s_k - p_k} = c$ for all $k$. 
Here, $c = \sqrt{\delta / 2^N}$. 

In the circumstance, $P(f(\bi{x})|\bi{x})$ of a classical near-solution machine is necessarily smaller than $1$ depending on $\delta$. 
On the other hand, if we choose the optimal phases $\phi_k$, $P(f(\bi{x})|\bi{x})$ can be 1 without any $\delta$-dependence in quantum machine. 
To show this, let us first write the conditional probability $P(f(\bi{x})|\bi{x})$ in \Eref{EQ:FreducedGeneral} as 
\begin{equation}
	P(f(\bi{x})|\bi{x}) = \abs{ \bra{f(\bi{x})} \left(\prod_{k \in A_\bi{x}} \hat{G}_k \right)\ket{0} }^2,
	\label{EQ:Pgeneral} %EQ:Pgeneral
\end{equation}
where $A_\bi{x}$ is the index set whose elements are indices of the actually applied operators conditioned on the input $\bi{x}=\{x_1,x_2,\dots,x_N\}$. 
For example, if $\bi{x}=1$ (i.e. $\{1,0,0\dots,0\}$ in the binary representation), then we have $A_\bi{x}=\{0,1\}$ because $G_0$ is always applied independently with the input, and the input signal $x_1=1$ activates $G_1$ (See \Fref{FIG:machine}). Thus, $\prod_{k \in A_1}\hat{G}_k = \hat{G}_1\hat{G}_0$. 
Based on the above description, we can generalize the calculations as
\begin{equation}
	\left\{
		\begin{array}{ll}
			\prod_{k \in A_1}\hat{G}_k = \hat{G}_1\hat{G}_0 & \text{for}~\bi{x}=1, \\
			\prod_{k \in A_2}\hat{G}_k = \hat{G}_2\hat{G}_0 & \text{for}~\bi{x}=2, \\
			\prod_{k \in A_3}\hat{G}_k = \hat{G}_3\hat{G}_2\hat{G}_1\hat{G}_0 & \text{for}~\bi{x}=3, \\
			~~~~~~~\vdots
		\end{array}
	\right.
	\label{EQ:prodUnitaries} %EQ:prodUnitaries
\end{equation}
Here, \Eref{EQ:Pgeneral} becomes to 1 when $c=0$ or equivalently $d(\mathrm{Q},\mathrm{S})=0$, because it is nothing but the solution machine. 
The basic idea is to find a condition that all terms of $c$ vanish even though $c$ is non-zero, i.e. the near-solution machine. 
Therefore, $P(f(\bi{x})|\bi{x})$ of the near-solution machine is mathematically equal to that of the solution machine. 
To do this, we consider the product of arbitrary two unitaries $\hat{G}_k\hat{G}_l$ ($k \neq l$), as
\begin{equation} 
	\left(
		\begin{array}{cc}
			\sqrt{p_k} 				& \rme^{\rmi \phi_k} \sqrt{q_k} \\
			\rme^{-\rmi \phi_k} \sqrt{q_k} 	& -\sqrt{p_k} \\
		\end{array}
	\right)
	\left(
		\begin{array}{cc}
			\sqrt{p_l} 				& \rme^{\rmi \phi_l} \sqrt{q_l} \\
			\rme^{-\rmi \phi_l} \sqrt{q_l} 	& -\sqrt{p_l} \\
		\end{array}
	\right).
\end{equation}
If we consider the near-solution machine, we can let $p_{k(l)} = \abs{s_{k(l)} - c}$ and $q_{k(l)} = 1-p_{k(l)}$. We then calculate $\hat{G}_k\hat{G}_l$, for the given $s_k, s_l$ in $\mathrm{S}$, as
\begin{equation} \fl
	\begin{array}{ll}
		\hat{G}_k\hat{G}_l = \left(
		\begin{array}{cc}
			\rme^{\rmi\Delta}(1 + \rme^{-\rmi\Delta}c\Lambda_-) &
			g(c)\rme^{\rmi\phi_l}\Lambda_- \\
			g(c)\rme^{-\rmi\phi_k}\Lambda_- &
			\rme^{-\rmi\Delta}(1 - c\Lambda_-) \\
		\end{array}
		\right) &~\text{if } s_k=0, s_l=0 ,\\
		\\
		\hat{G}_k\hat{G}_l = \left(
		\begin{array}{cc}
			g(c)\Lambda_+ &
			\rme^{\rmi\phi_l}(1 - c\Lambda_+) \\
			-\rme^{-\rmi\phi_l}(1 + c\rme^{-\rmi\Delta}\Lambda_+) &
			g(c)\rme^{-\rmi\Delta}\Lambda_+ \\
		\end{array}
		\right) &~\text{if } s_k=1, s_l=0, \\
		\\
		\hat{G}_k\hat{G}_l = \left(
		\begin{array}{cc}
			g(c)\Lambda_+ &
			-\rme^{\rmi\phi_k}(1 - c\rme^{-\rmi\Delta}\Lambda_+) \\
			\rme^{-\rmi\phi_k}(1 - c\Lambda_+) &
			g(c)\rme^{-\rmi\Delta}\Lambda_+ \\
		\end{array}
		\right) &~\text{if } s_k=0, s_l=1, \\
		\\
		\hat{G}_k\hat{G}_l = \left(
		\begin{array}{cc}
			1 - c\Lambda_- &
			-g(c)\rme^{\rmi\phi_k}\Lambda_- \\
			g(c)\rme^{-\rmi\phi_k}\Lambda_- &
			1 - c\rme^{\rmi\Delta}\Lambda_- \\
		\end{array}
		\right) &~\text{if } s_k=1, s_l=1,
	\end{array}
	\label{EQ:GkGl} %EQ:GkGl
\end{equation}
where $\Lambda_\pm = 1 \pm \rme^{\rmi\Delta}$, $\Delta = \phi_k - \phi_l$, and $g(c) = \sqrt{c-c^2}$. In calculating \Eref{EQ:GkGl}, we assumed a deterministic target, i.e. $s_{k(l)}$ is to be either $0$ or $1$, as it is usual in most tasks (but not necessarily). Here, the important thing is that we can vanish the term associated with $c$, by letting 
\begin{equation}
	\Lambda_\pm = 0, ~\text{or equivalently},~
		\left\{
			\begin{array}{cc}
				\phi_k = \phi_l	&~~\text{if } s_l = s_k, \\
				\phi_k = \phi_l + \pi	&~~\text{if } s_l \neq s_k.
			\end{array}
		\right.
	\label{EQ:phaseCondition} %EQ:phaseCondition
\end{equation}
The above condition in \Eref{EQ:phaseCondition} can be applied for all $k \neq l$. Thus, we provide here a generalized condition as
\begin{equation}
	\phi_k = 
	\left\{
		\begin{array}{cc}
			0	&~~\text{if } s_k = 0, \\
			\pi	&~~\text{if } s_k = 1,
		\end{array}
	\right.
	\label{EQ:optimizedPhase2} %EQ:optimizedPhase2
\end{equation}
This is the optimal phase condition, as in \Eref{EQ:optimizedPhase}. We can check that this condition gives the maximum task-fidelity with $P(f(\bi{x})|\bi{x})=1$ (for all $\bi{x} \neq 0$).

%%%%%%%%%%%%%%%%%%%%%%%%%%%%%%%%%%%%%%%%%%%%%%%%%%%%%%%%%%%%
\section{A practical version of quantum machine}\label{appendix_b}
%%%%%%%%%%%%%%%%%%%%%%%%%%%%%%%%%%%%%%%%%%%%%%%%%%%%%%%%%%%%
The speedup introduced in this paper is enabled when quantum machine uses suitable phases. 
Accordingly, the suitable phases are pre-required for the fast learning. 
In a practical manner, the learning-time has to include complexity to get suitable phases which is not so easy to get. 
We introduce a practical quantum machine that does not require the effort in finding an optimal phases. 
To this end, we modify the unitary $\hat{G}_k$ in \Eref{EQ:quantumUnitary} by setting all the phases $\phi_k$ to $\pi p_k$, i.e., $\hat{G}_k$ is written as
\begin{equation}
	\hat{G}_k =
		\left(
			\begin{array}{cc}
				\sqrt{p_k} 				& \rme^{\rmi \pi p_k} \sqrt{1-p_k} \\
				\rme^{-\rmi \pi p_k} \sqrt{1-p_k} 	& -\sqrt{p_k} \\
			\end{array}
		\right),
\label{EQ:quantumUnitary2} %EQ:quantumUnitary2
\end{equation}
such that phases $\pi p_k$ are getting closer to the optimized phases $\pi s_k$ as the machine approaches to the solution point in parameter space during the learning, since the optimized phase condition is given by \Eref{EQ:optimizedPhase}. 
Thus, this guarantees wider acceptable regions than classical machine for any learning target. 

%Fig5_acceptableRegion.pdf, Fig5_learning_time.pdf
\begin{figure}[tp]
	\centering
	\subfigure[acceptable region]
		{\includegraphics[width=2.4in]{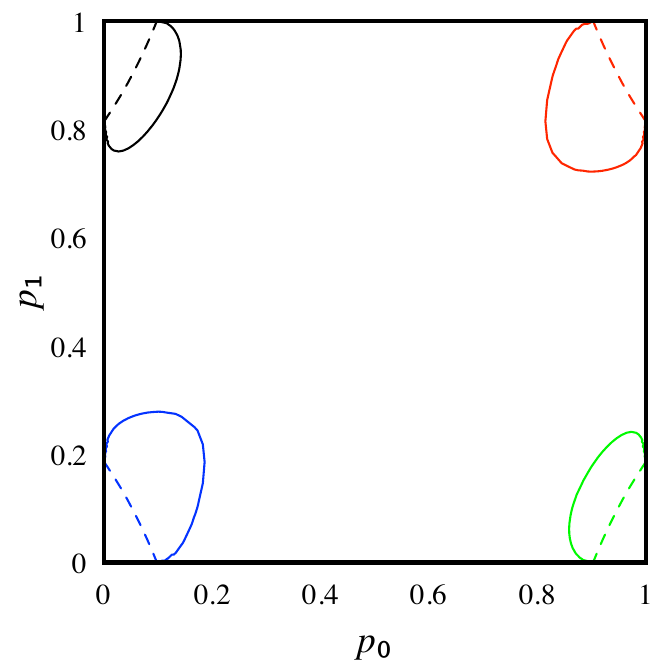}}
	\subfigure[learning time]
		{\includegraphics[width=3.225in]{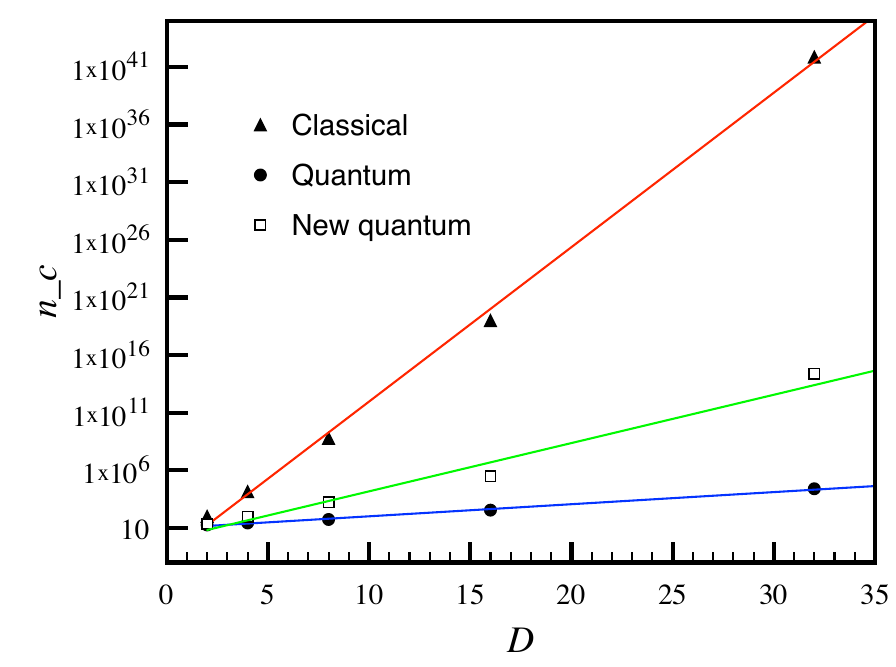}}
	\caption{
		(a) We depict the boundary of the acceptable regions of $1$-bit learning machine with respect to all possible target function, as in \Tref{TAB:1bitBoolean}: $f_1$(red), $f_2$(black), $f_3$(green) and $f_4$(blue). 
		We set $\epsilon_t = 0.05$. It is directly seen that the quantum (solid line) has larger acceptable regions than the classical (dotted line) for all cases. 
		(b) The learning-time, $n_{c}$, with dimension $D$ of parameter space. 
		Data for classical(red) and quantum(blue) machines are exactly same to \Fref{FIG:result_RS}. 
		Data for new quantum machine(green line) are well fitted to $\ln{n_c}=\alpha D + \beta$, with the fitting parameters $\alpha\simeq 0.985$ and $\beta\simeq -0.200$.
	}
	\label{FIG:PhaseParameterSync} %FIG:PhaseParameterSync
\end{figure}

\Fref{FIG:PhaseParameterSync} (a) shows that the practical quantum machine has wider acceptable regions than classical machine for all $1$-bit Boolean targets. 
Inside areas of solid and dashed lines represent acceptable regions of the practical quantum machine and classical machine, respectively. 
This supports that the practical quantum machine always learns faster than classical machine, while original quantum machine depends on target function.

We then obtain learning-time of the practical quantum machine in \Fref{FIG:PhaseParameterSync} (b). 
The data are also well fitted to $\ln{n_c} = \alpha D + \beta$, with the fitting parameters $\alpha \simeq 0.985 \pm 0.101$ and $\beta \simeq -0.200 \pm 1.662$. 
Thus, $n_c \sim O(\rme^{0.985 D})$ in the practical quantum machine, whereas $n_c \simeq O(\rme^{3.065 D})$ in the classical machine (See \Eref{EQ:fittingParameter}). 
The result shows that a considerable learning speedup is still achieved in this practical quantum machine, even though it takes little bit more time compared to original one available with the optimal relative phases ($n_c \sim O(\rme^{0.238 D})$).

%%%%%%%%%%%%%%%%%%%%%%%%%%%%%%%%%%%%%%%%%%%%%%%%%%%%%%%%%%%%
\section*{References}
%\bibliography{references}
%\bibliographystyle{iop}

%%%%%%%%%%%%%%%%%%%%%%%%%%%%%%%%%%%%%%%%%%%%%%%%%%%%%%%%%%%%

\end{document}